\documentclass[10pt,journal,letterpaper,compsoc]{IEEEtran}
\usepackage{ifpdf}

\ifCLASSOPTIONcompsoc
   \usepackage[nocompress]{cite}
\else
   \usepackage{cite}
\fi
\ifCLASSINFOpdf
   \usepackage[pdftex]{graphicx}
\else
   \usepackage[dvips]{graphicx}
\fi
\usepackage[cmex10]{amsmath}
\usepackage{algorithmic,algorithm}

\usepackage{array}

\usepackage{mdwmath}
\usepackage{mdwtab}

\usepackage{eqparbox}
\newtheorem{axiom}{Axiom}

\begin{document}
%
\title{Originator usage control with business process slicing}
%
%
%
%

\author{Ziyi~Su,
        and Fr\'ed\'erique~Biennier
\IEEEcompsocitemizethanks{
\IEEEcompsocthanksitem F. Biennier is full professor of Lab. LIRIS CNRS, Department d'informatique, INSA Lyon, France, 69100.\protect\\
E-mail: see http://liris.cnrs.fr/frederique.biennier/
\IEEEcompsocthanksitem Z. Su is doctoral student of Lab. LIRIS CNRS, Department d'informatique, INSA Lyon.
}
\thanks{This work was partially supported by DGCIS project 'process2.0' and ANR project 'semeuse'. }}

%
%

\markboth{(IEEE format)}%
{Shell \MakeLowercase{\textit{et al.}}: Bare Demo of IEEEtran.cls for Computer Society Journals}
%



\IEEEcompsoctitleabstractindextext{%
\begin{abstract}
Originator Control allows information providers to define the information re-dissemination condition. Combined with usage control policy, fine-grained 'downstream usage control' can be achieved, which specifies what attributes the downstream consumers should have and how data is used. This paper discusses originator usage control, paying particular attention to enterprise-level dynamic business federations. Rather than 'pre-defining' the information re-dissemination paths, our \emph{business process slicing} method 'capture' the asset derivation pattern, allowing to maintain originators' policies during the full lifecycle of assets in a collaborative context.
First, we propose Service Call Graph (SCG), based on extending the System Dependency Graph, to describe dependencies among partners. When SCG (and corresponding 'service call tuple' list) is built for a business process, it is analyzed to group partners into sub-contexts, according to their dependency relations. Originator usage control can be achieved focusing on each sub-context, by examining downstream consumers' security profiles with upstream asset providers' policies.
Second, for analyzing SCG, we propose two 'slicing' strategies, namely 'asset-based' and 'request-based' slicing, to deal with the scenarios of both 'pre-processing' a business process scripts and 'on-the-fly' analyzing service compositions.
Last, our implementation work involves a 'context manager' service for processing business processes defined in WS-BPEL. It can be composed with our former proposed policy negotiation and aggregation services to provide policy-based end-to-end security management.
We also make experiments based on processing the sample processes that come with 'WS-BPEL2.0' specification.

\end{abstract}

}

\maketitle

\IEEEdisplaynotcompsoctitleabstractindextext

%
\IEEEpeerreviewmaketitle

\section{introduction}\label{intro}

In collaborative systems, participants share digital assets, including computing capability (e.g. Web Service) and information (e.g. data), in order to produce a final artifacts (e.g. composed service, new information generated from data aggregation). 
As assets are shared beyond ownership boundary, risk of Intellectual Property infringement  (e.g. circumventing of trade secret, or even leakage to a competitor) associated to 'loss of governance' is the major barrier for moving toward collaborative business model \cite{Linda2010} \cite{kagal10Jul} \cite{Daniele2009}. There security requirement is brought to an end-to-end scale: to ensure the protection of corporate patrimony value during it full lifecycle in the collaborative business process, covering the creation, derivation and destruction stages, paying particular attention on how the asset is used by partners. Overcoming this barrier relies on a comprehensive method that can make the 'pre-decision' about selecting partners based on historical comportment, as well as can continuously regulate the partners behaviors on the fly in the collaborative business process.


Such requirement leads to the pursuit of Originator Control (ORCON) with Usage control policies \cite{Park:2002:OCU:863632.883494} (or Downstream Usage control \cite{LGF2010rohtua}), where the goal is to require recipients to gain originator's approval for re-dissemination an originally digital asset or a new digital asset that includes it \cite{Park:2002:OCU:863632.883494}.
Recent progresses in 'usage control' policies \cite{Ni2011} \cite{Karat:2009:PFS:1850636.1850640}  \cite{Zhang2008a} pave the bedrock for fine-grained control of 'due usage' actions and 'obligations' on digital assets.
For now a stage, Originator Usage Control policies are mainly built around the strategy of close-coupling the control of assets dissemination paths and the control of 'due usage' in each hop \cite{LGF2010rohtua} \cite{Park:2002:OCU:863632.883494}. This strategy suit well scenarios where providers want to have direct control on the exact dissemination steps. But whan applied to more general business federation scenarios, the close-coupling strategy can be cumbrous. For now, collaborative business processes are achieved mainly by 'on-the-fly' service composition or pre-defined orchestration (e.g. with WS-BPEL). In either case, the originator's end-to-end security requirements concern choosing the partners that possess eligible 'Quality-of-protections' for its asset. Once the originators' security criteria are met, the decision of choosing which partners (i.e. the exact paths of asset dissemination) is more related to the business logic. Then, in the perspective of security engineering for the collaborative context as a whole, ensuring that all the providers' security criteria are maintained during the full lifecycle of their assets, is the essence. This involves analyzing the collaborative business process to track the dissemination paths of every asset, the co-effect of their providers' policies calculated when assets merge, the consumer's security profile (and 'usage' activities) checked when an assets is consumed.

In former works we have proposed a collaborative usage control model that deduces the co-effect of policies \cite{ZSuFB2012a} and an implementation architecture supporting 'usage control' enforcement \cite{Biennier2010}. It's our intention in this paper to complete our end-to-end security management framework, by developing a method for analyzing complex collaborative context and applying collaborative usage policy to manage asset sharing activities. The basic thought is that security foundation for a successful collaborative process is that each originator's policy is fulfilled during the whole business process.

For holistic view, we briefly introduce our analysis of the asset derivation patterns and sub-context modes involved in collaborative business processes. Security management can be done in the scope of each 'sub-context'.

The first contribution of this paper is a 'Service Call Graph' (SCG) we developed based on modifying and extending System Dependency Graph (SDG),  for representing partners' interactions in collaborative business processes. We also propose a data structure 'service call tuple' corresponding to the SCG for capturing 'dependencies' among partners.

Then We propose 'asset-based' and 'request-based' context slicing methods, for mining the 'asset (RoP) aggregation' and 'request (QoP) aggregation' from the 'service call tuple' list that represents a business process. Such aggregations decides the partitions of sub-contexts, where collaborative usage control policies can be applied.

We analyze the sub-context developments, using 'pre-processing' and 'on-the-fly processing' strategies, and describe how originator usage control is achieved by managing sub-context developments. 

We also implement a 'context manager' service for the context slicing task, paying particular attention to complex business process defined in WS-BPEL. This service can be deployed with other components of our 'collaborative usage control policy' enforcement system [], to provide originator usage control service. Experiment results of processing the sample business processes that come with 'WS-BPEL2.0 specification' are presented.
%
%
%

Section (\ref{context}) introduces the 'usage control' model we use as foundation for ORCON, and the Service Dependency Graph (SDG), based on which we develop a business process 'slicing' method. The difference between our approach and several representative ORCON works are discussed.
Section (\ref{Sub-context modes}) gives a general level description of the assets aggregation patterns usually involved in a business process.
Section (\ref{Sub-c slicing}) introduces the business process analysis method, led by a motivating use case.
Section (\ref{GCbp}) discusses the capability of our analysis method with a more comprehensive use case.
Implementation and experiment results are introduced in section (\ref{Implmt}).

\section{Background and related works}\label{context}
Originator Control (ORCON) aims at enforcing providers' policies during the full lifecycle of assets put to a collaborative context. Combining with 'usage control' policies enhance its expressivity w.r.t. defining the 'consumption activities' and the access condition.

\subsection{Originator control (ORCON)}\label{Fitting federation}
ORCON was brought forward to restrict the distribution of documents in paper world \cite{ORCON0}, where the approval of originator is always required. Traditional ORCON solutions in cyberspace aim at automate ORCON \cite{Sandhu1992} \cite{McCollum1990}, using some form of non-discretionary access control list  \cite{Park:2002:OCU:863632.883494}. Recent works combine ORCON with usage control, greatly enhanced its expressivity \cite{Park:2002:OCU:863632.883494} \cite{LGF2010rohtua}. But these works center on allowing originators to control the exact path their asset are re-distributed, therefore is more related to the 'information dissemination' control, which are mostly encountered in security research of social network or pub/sub systems. Whereas in enterprise level federations, the assets dissemination path is more closely related to the business logic. In other words, the originators' security concern is that their assets must only be passed to partners with eligible security attributes (encrypted communication channel, secured platform, organizational security management convention, etc.) and allowed 'usage' activities. As soon as these criteria is met, the exact chosen partners and asset sharing 'paths' are decided mainly according to 'common business goal' (e.g. functional requirements and QoS requirement in Web Service composition).  Our work focuses on the enterprise level business federation scenarios and therefore is different from the former originator usage control works in that we identify the partners that the originator will share assets with, allowing to exam their security profiles with the originator's requirements, based on collaborative usage control policies.

\subsection{Collaborative usage control policy}\label{OCNplc}
A basic usage control policy model can be described by the following factors:
\begin{equation}
UCON= (S, O, Ct, Rt, Ob, Rn, P	)
\label{scheme}
\end{equation}
where
\begin{itemize}
\setlength{\itemsep}{1pt}
\setlength{\parskip}{0pt}
\setlength{\parsep}{0pt}
\item '\(S\)' (Subject) is the party that can get a Right on the asset. It is specified by a set of 'Subject Attributes' (\(SAT\)).
\item '\(O\)' (Object) is the asset to be protected by the policy rule. It is specified by a set of 'Object Attributes' (\(OAT\)).
\item '\(Ct\)' (Context) is the collaboration context that is associated to the status of the system infrastructure, the environment and the business federation. It is specified by the set of 'Context Attribute'(\(CNAT\))
\item '\(Rt\)' (Right) is the Operation upon the asset defined by 'Sh' that the Subject is allowed to achieve.
\item '\(Ob\)' (Obligation) is the obligation that must be fulfilled by the Subject when it gets the Right.
\item '\(Rn\)' (Restriction) is the attribute (from 'Context' set) associated to 'Rt' or 'Ob' to further confine in what circumstance they are carried out.
\item '\(P\)' (Policy)is the usage control policy definition. It maps predicates on a set of attributes identifying 'Subject', 'Object', 'Context' to the predicates on a set of attributes identifying 'Right' or 'Obligation'.
\end{itemize}
\label{schemedef}

Some works lay the foundation of 'usage control' system. In the $UCON_{ABC}$ model \cite{Zhang2008a} \cite{Park2004} \cite{Zhang-2006-p180-189}, access decision is based on not only role (as in RBAC) or identity (as in MAC and DAC), but on diverse attributes of subjects and objects. With the grant and exertion of a 'usage' right, multiple consumption actions (e.g. playing several songs) can happen, during this process, the attributes of the object (e.g. the amount of the 'not used' objects) and subject (e.g. balance in her/his account) are constantly changing. The authors defined models for 'attribute-update' actions, which capture the semantic of this process. A formalization dedicated to the $UCON_{ABC}$ model has been introduced \cite{Zhang-2005-p351-387} \cite{Zhang-2004-p1-10} using Lamport's Temporal Logic of Action (TLA). It describes the temporal constrain between the action in Authorization (granting of right), Obligation and attribute update that results from the exercising of granted rights or imposed obligations. Manuel Hilty et al. proposed a usage control policy language \cite{Hilty-2008-p531-546} based on the analysis of usage control requirements and existing control mechanisms \cite{Hilty2006} \cite{Pretschner2006}. Their work gave a taxonomy on usage actions, as well as the common attributes of subject, object and environment, e.g. time, cardinality ('how many times' an action can be performed), events, purpose of usage, etc.

In collaborative context, when assets merge, their providers' policies should also aggregate, to achieve ORCON. One fundamental issue is to set the aggregation method so that the resulting policy correctly interprets the original goal of the providers' policies. In our former work \cite{ZSuFB2012a}, we built a collaborative usage control scheme by introducing several factors:
\begin{itemize}
\setlength{\itemsep}{1pt}
\setlength{\parskip}{0pt}
\setlength{\parsep}{0pt}
\item '\(Sh\)' (Stakeholder) is the owner of the rule, and is the owner or co\_{}owner of the assets related to the rule.
\item '\(lc\)' (Lifecycle)  defines the lifecycle \cite{ZSuFB2011c} of a predicate, extending the effect of the predicate to (indirect) partners corelated by the business federation, e.g. predicates tagged with $lc=dp$ take effect only between 'direct partner', effects of predicates tagged with $lc=eot$ are maintained till the 'end of transaction'.
\item '\(G\)' (Aggregation algorithm) is the algorithm used to 'combine' the individual policies from each 'Sakeholder'.
\end{itemize}
\label{schemedef}


Our policy scheme follows such design pattern that providers use '$RoP$' policies to express their 'requirements of protection' upon their assets.
Consumers use $QoP$ (analogous to the P3P \cite{Cranor2002} approach) to express their 'Quality of Protection', e.g. 'promises' about the protection they offer (they way they consume asset, their security attributes, etc.). When two assets merge, policies (the \(RoP\)s that are refined with '$lc$') upon them are aggregated. For the partners who will consumer the same set of assets, their $QoP$s are aggregated. The aggregation algorithm detect potential conflicts between the policies that will merge, in order to find inconsistent security attributes (between $QoP$s) or requirements (between $RoP$s).

In short, this policy use a strategy similar to the 'stick policy' method \cite{Mont-2003-p377-377} \cite{Chadwick-2008-p1-6} \cite{Bandhakavi-2006-p51-58}, to allow propagating providers policies to the artifact of collaborative context, when their assets are merged into it.
The advantage of using  scheme to manage collaborative context security is that security configuration can be done in a peer-to-peer manner: if every partners' security requirements are met, the condition (in security perspective) for context to be set is fulfilled. But for this, partners' assets sharing relations should be identified. For a holistic view, next section briefly introduces our former work on analysis of the sub-context modes \cite{zsfbAPMS11} incurred by asset sharing activities.

\subsection{Sub-context modes}\label{Sub-context modes}
Basically, only partners corelated by (both direct or indirect) assets exchanging a deemed in on sub-context. For convenience of discussion, we denote the assets provided by partners as 'Original Assets' ('O-Assets' for short) and the artifacts of collaborative work (aggregating several O-Assets) as 'Collaboration Assets' ('C-Assets' for short).

Security management with the asset aggregation viewpoint are based on the following \textbf{principles}:
\begin{itemize}
\setlength{\itemsep}{1pt}
\setlength{\parskip}{0pt}
\setlength{\parsep}{0pt}
\item [\emph{\textbf{1}}.]All participants having the \textbf{same 'Rights'} upon the \textbf{same C-Asset(s)} are gathered in one \emph{sub-context}.
\item [\emph{\textbf{2}}.]Each sub-context has its \textbf{\emph{Context Security Policy ('CSP' for short)}}, which has two parts: \textbf{$RoP_{CSP}$} includes the providers' \(RoP\)s and represents the \textbf{providers' security requirements on the C-Asse}t; \textbf{$QoP_{CSP}$} includes the consumers' \(QoP\)s and represents the \textbf{the sub-context's security profile} for future assets providers.
\item [\emph{\textbf{3}}.]A participant can belong to \textbf{more than one sub-context at the same time}. It must follow the CSPs of all the sub-contexts it belongs to.
\end {itemize}

\begin{figure}[htbp]
\centering
\includegraphics[width=0.50\textwidth]{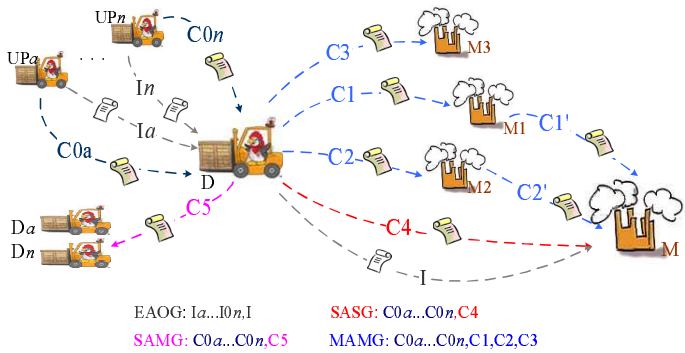}
\caption[Sub-contexts patterns in business federation]{Sub-contexts patterns in business federation} \label{fig:graph505}
\end{figure}

The central issue is to analyze the provider-consumer relation among partners. Note that a participant in the collaborative context can act as both provider and consumer, according to the business logic, which make the analysis tricky. We use a sample supply chain scenario (see figure \ref{fig:graph505}) to presents the sub-context modes one can encounter when analyze a collaborative context. 

We see 4 basic sub-context partition modes. 

\begin{itemize}
\setlength{\itemsep}{1pt}
\setlength{\parskip}{0pt}
\setlength{\parsep}{0pt}
\item 'Each Asset One Group (EAOG)' mode:\\
When \textbf{all providers can distinguish their own asset (O-Assets) in the artifact of business collaboration (C-Asset)}, there is no asset aggregation, each partners works in a separate sub-context. 
An example (see 'EAOG'in \ref{fig:graph505}) is when a down-stream provider (D) receives inventory information '$I_a$...$I_n$' from upstream providers (UP) and concatenates them in one XML file '$I$', each of them as a separate node. The manufacturer (M) reading one nodes must follow the due policy of UP.
\item 'Single-Asset Single-Group (SASG)' mode:\\
If the O-Assets in C-Asset are \textbf{not identifiable}, their providers are deemed as in one \textbf{single group}, given that they define the same rights. 
An example (see 'SASG'in \ref{fig:graph505}) is when production information '$C_{0a}$...$C_{0n}$' are merged by 'D' to generate a global scheduling '$C_4$'. 'M' reads must fulfill the policies of all the 'UPs' in order to read $C_4$.  
\item 'Single-Asset Multi-Group (SAMG)' mode:\\
If the O-Assets in C-Asset are \textbf{not identifiable}, by their providers have define multiple rights, \textbf{each right forms a group}.

An example (see 'SAMG'in \ref{fig:graph505}) is when the 'UPs' all define multiple rights (e.g. 'read' and 'disseminate'). Then their policies co-define multiple rights on $C_5$ has multiple rights. Consumers holding different Rights upon the same $C_5$ are deemed as in different sub-contexts.
\item 'Multi-Asset Multi-Group (MAMG)'mode:\\
If there are multiple C-Assets and an O-Asset can belong to more than one C-Asset at the same time. We should identify which O-Assets disseminate in which C-Assets, in order to examine, firstly, whether the providers' policies of aggregated assets are compatible and, secondly, whether the security profiles of the consumers for a C-Asset fulfill the policies (coming from the policies of O-Assets providers) on that C-Asset.

\end {itemize}

%

These 4 patterns generally exist in many collaborative contexts, as long as the issue of information asset protection and consumption exists. Imagining changing above supply chain scenario to others, e.g. switching the materials providers as Cloud providers or Service providers, the asset exchange pattern still fall in the 4 modes.


The CSP aggregation process involves detecting potential conflicts, with a algorithm we proposed before \cite{ZSuFB2012a} 

The center issue in this paper is to develop a method that analyzes the collaborative business process to partition sub-contexts and allocate partners according to asset sharing patterns. We develop a method for this task, borrowing (and modifying) the method used for programm slicing with System Dependency Graph (SDG).

\subsection{System dependency graph}\label{Service dependency}
In business federation, assets transfer across organization boundaries, possibly merging with other assets. In order to give a life long protection to an asset, it's necessary to capture the asset derivation relations and track the asset in the artifacts of business process. This issue is similar in its nature to the 'Program Slicing' \cite{grammatec} \cite{Zhao03systemdependence} task based on System Dependency Graph (SDG) \cite{grammatec} \cite{DBLPGuDDXM08}. Program slicing asks about which statements influence (backward slice), or is influenced by (forward slice), the current statement under exam. Whereas collaborative process analysis asks about which processes (functionalities provided by a partner can be seen as a process, e.g. implemented with a Web Service) influence which processes, therefore tracing asset exchange and derivations.

We use a similar approach to 'slice' a collaborative context into 'sub-contexts'. Each 'sub-context' confine a scope of partners interrelated by assets exchanges (in other words, partners in different 'sub-context' don't exchange asset, although they are in the same collaborative process). We firstly give an overview of the 'sub-context' modes we may encounter when analyzing a collaborative context, before introducing the analysis method.

%

%
%
%
%
In the following sections, we fist discuss the analysis of a collaborative context with SASG mode, guided by a simple use case, in order to introduce the basic thoughts underlaying out analysis method. Then we use a full-fledged use case to describe how our method is used to deal with complex business process that is in MAMG mode.

\section{Context analysis based on 'context slicing'}\label{Sub-c slicing}
\begin{figure}[htbp]
\centering
\includegraphics[width=0.45\textwidth]{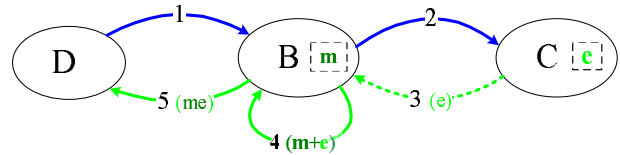}
\caption[Service composition represented with SCG]{Service composition represented with SCG} \label{fig:graph334}
\end{figure}

We use a simple use case (see figure \ref{fig:graph334}) of Web Service composition to facilitate our discussions.

\textbf{Use case 1}: An assurance association 'Deirect assure'(D) consults 'medical information' ($m$) from 'Bonetat clinique'(B). Part of the information, 'cardiac exam' ($e$), is taken from a medical examination laboratory 'Cardis health'(C). The business process includes the following steps:

(1) D contacts B, requiring $m$;

(2) B contacts C, requiring $e$, in order to reply D, if success,

(3) C sends $e$ to B;

(4) B merges $e$ with $m$; if success,

(5) B answers to D.
%


As B and C are asset (information) providers in this use case , ORCON means that their policies should be respected during the whole lifecycle of their assets. This involves answering two questions:
\begin{itemize}
\setlength{\itemsep}{1pt}
\setlength{\parskip}{0pt}
\setlength{\parsep}{0pt}
\item (1) Which partners will access an asset? A question of this type for 'use case 1' is "The 'Cardiac exam info'  provided by C will be accessed by B or D, or both of them?"
\item (2) Which assets will be accessed a party? As a simple illustration, in 'use case 1' a question of this type is "Whether D will access the assets provided by B and C, either directly or indirectly".
\end{itemize}

While both of these questions can be answered intuitionally for 'use case 1', the pondering procedure reflects the goal and method of context slicing. Question (1) is related to QoP aggregation among partners. Question (2) is linked to RoP aggregation. The goal is to enable the 'down-stream usage control' \cite{LGF2010rohtua}, so that indirect consumers should follow the policies of the O-Assets involved in the C-Asset they access. The method we use is analogous to the 'Program Slicing' \cite{grammatec} \cite{Zhao03systemdependence} based on System Dependency Graph (SDG) \cite{grammatec} \cite{DBLPGuDDXM08}.


\subsection {Service call graph} \label {SCG}
A participant in a collaborative contexts is analogous to a 'procedures' in a SDG: receives calling information and produce results. We use $P_i \stackrel{c}{\longleftarrow} P_j$ to denote that a party $P_i$ depends on another party $P_j$ with 'control dependency': whether $P_i$ will be activated depends on $P_j$.  $P_i \stackrel{d}{\longleftarrow} P_j$ denotes that a party $P_i$ depends on another party $P_j$ with 'data dependency': data provided by $P_j$ are involved in data produced by $P_i$. We propose a data structure 'Service Call Graph' (SCG) based on extensions of SDG to represent partner interactions in the collaboration context. These extensions can be illustrated with 'use case 1' (figure \ref{fig:graph334} is a SCG of 'use case 1'):

\begin{itemize}
\setlength{\itemsep}{1pt}
\setlength{\parskip}{0pt}
\setlength{\parsep}{0pt}
\item
The fist extension is that '\emph{\textbf{data dependency}}' in our SCG is differentiated as two types: an '\emph{\textbf{aggregation dependency}}' means $P_i$ involves data of $P_j$ (the same as SDG), a '\emph{\textbf{non-aggregation dependency}}' denoting that data produced by $P_i$ does not involve data from $P_j$ (an extension of SDG). For example, in the SCG presented in figure (\ref{fig:graph334}), the \emph{\textbf{blue edges}} (step 1 and 2) represent '\emph{\textbf{control dependency}}'. The \emph{\textbf{green edges}} (steps 3, 4 and 5) represent '\emph{\textbf{data dependency}}', where The \emph{\textbf{solid green lines }}(edge 4 and 5) means that the output data (response) \emph{\textbf{includes}} information from the input data (aggregation dependency). The \emph{\textbf{dashed green line}} (edge 3) means that the output data \emph{\textbf{does not include}} information from the input data (non-aggregation dependency).

%
\item
The second extension is that the \emph{\textbf{assets carried by the message}} exchanges are \emph{\textbf{attached directly to the edges}}  in SCG (see edges 3, 4 and 5 in figure \ref{fig:graph334}).
\item
The last extension is that we represent \emph{\textbf{failed}} interactions (due to negative result of policy negotiation) with \emph{\textbf{dashed blue lines}}. A more comprehensive use case presented in section \ref{GCbp} will give such examples.
\end{itemize}

In order to capture assets derivation pattern, we define 'indirect dependency', based on partner service call in a business collaboration: $\forall P_i, P_j, P_k, \forall \alpha \in \{c,d \}$ where $P_i$, $P_j$ and $P_k$ are partners in a collaboration, $c$ and $d$ are 'control dependency' and 'data dependency' relations respectively, $P_i$ is 'indirectly dependent' on $P_k$, if '$P_i \stackrel{\alpha}{\longleftarrow} P_j \wedge P_j \stackrel{\alpha}{\longleftarrow} P_k$'.

There are two types of indirect dependency.
'\emph{Indirect \textbf{data} dependency}' is the situation where each relation in a dependency chain is 'data dependency'. We sum it up as an axiom:

\begin{axiom}[Indirect data dependency]
$\forall P_i, P_j$, $P_k$: $P_i \stackrel{d}{\longleftarrow} P_j \wedge P_j \stackrel{d}{\longleftarrow} P_k \Rightarrow P_i \stackrel{d}{\longleftarrow} P_k$
\label{indatadpdnc}
\end{axiom}

For example, in 'use case 1', whether D gets the results or not depends on the response of B. B's response in turn depends on response from C.

'\emph{Indirect \textbf{control} dependency}' is the situation where 'control dependency' relation exists in the dependency chain:

\begin{axiom}[Indirect control dependency]

$\forall P_i, P_j, P_k, \forall \alpha \in \{c,d \}$: $(P_i \stackrel{c}{\longleftarrow} P_j \wedge P_j \stackrel{\alpha}{\longleftarrow} P_j)\vee(P_i \stackrel{\alpha}{\longleftarrow} P_j \wedge P_j \stackrel{c}{\longleftarrow} P_j) \Rightarrow P_i \stackrel{c}{\longleftarrow} P_k$
\label{incntldpdnc}
\end{axiom}


As an example for indirect control dependency, in 'use case 1', whether C will be called depends on B. Whether B will be called in turn depends on D. So C is indirectly control dependent on D.

We can see the slight difference between axiom \ref{indatadpdnc} and axiom \ref{incntldpdnc}: Data dependency is transitive only when the edges in the dependency chain are all associated to 'data dependency', whereas when control dependency exists in a dependency chain, it propagates 'control dependency' to the chain.

When analyzing complex business process, e.g. those defined by WS-BPEL, one must consider the impact of 'variables', which are used to carry information inside the process. As information carried by 'variables' are eventually exchanged between partners, the information exchanges between 'variables' (e.g. through 'value assignment') also lead to assets derivation.

These variables can be complex data type (e.g. defined by XML schema). In this case, if a part of a variable is valued-assigned to a part of another variable (see the 'sample process' in WS-BPEL specification \cite{WSBPEL2007}), the later variable is 'data dependent' on the former one. Thus we have the following axiom:

\begin{axiom}[Direct data dependency between variables]

$\forall c_m \in P_i, c_n\in P_j$, $c_m \stackrel{d}{\longleftarrow} c_n \Rightarrow P_i \stackrel{d}{\longleftarrow} P_j$
\label{prmtdpdnc}
\end{axiom} where $P_i$ and $P_j$ stand for 'variables'. '$c_m$' is part of $P_i$. '$c_n$' is part of $P_j$.

There are only 'data dependency' relations between variables, as the only form of interactions between variable is data exchange. Therefore the conditions leading to 'indirect data dependency' between variables can be described by axiom \ref{indatadpdnc}. In the following discussion about dependency relation, we don't need to differentiate 'variables' from 'partners', as we can see that dependency relations for 'partners' and for 'variables' can be described by the same set of axioms.

\subsection {Service call tuple} \label {SCT}
We use a tuple $<P_i\stackrel{t}{\longleftrightarrow}P_j, \Delta>$ to denote the service call from $P_i$ to $P_j$, $\Delta$ being the exchanged asset. We can have the following basic types of service call tuple:
\begin{itemize}
\setlength{\itemsep}{1pt}
\setlength{\parskip}{0pt}
\setlength{\parsep}{0pt}
\item $<P_i\stackrel{c}{\longrightarrow}P_j>$ denotes that $P_i$ calls $P_j$ with a message carrying no asset.
\item $<P_i\stackrel{c}{\longleftarrow}P_j>$ denotes that $P_i$ receives a message from $P_j$ that carries no asset.

An example of these two types of service call is when a mail agent queries a mail service for whether a mail is sent, and receives confirmation from the server. In such case the call message and the response message are deemed as not carrying any asset (i.e. information needing protection). We can see that whether a message carries asset or not depends on the straining criteria of security in a specific application context.
\item $<P_i\stackrel{d}{\longrightarrow}P_j, \Delta_i>$ denotes that $P_i$ calls $P_j$, by sending asset $\Delta_i$.
\item $<P_i\stackrel{d}{\longleftarrow}P_j, \Delta_o>$ denotes that $P_i$ receives a response from $P_j$ that carries asset $\Delta_o$.
\item $<P_i\stackrel{\alpha}{\longleftrightarrow}P_j, \Delta_{i}, \Delta_{o}>$ denotes that $P_i$ calls $P_j$, sending asset $\Delta_i$ and receiving response carrying asset $\Delta_o$, where $\Delta_o$ includes information from $\Delta_i$.
\item $<P_i\stackrel{\alpha}{\longleftrightarrow}P_j, \Delta_{i}, \Delta_{o},\not\subset>$ denotes that $P_i$ calls $P_j$, sending asset $\Delta_i$ and receiving response carrying asset $\Delta_o$, where $\Delta_o$ does not include information from $\Delta_i$.
\item $<P_i\stackrel{f}{\longleftrightarrow}P_j,\not\subset>$ denotes that the interaction between $P_i$ and $P_j$ failed, due to negative result of policy negotiation.
\end{itemize}

These tuples represent the edges of SCG. 
We can see that asset exchanges (and aggregations) occur with service calls.

\subsection {Assets aggregation} \label{Assagg}
Usually, with partners' interactions, assets derivations (basically, either 'merging' or 'splitting') happen. Therefore, recognizing assets derivation relations involves firstly formalizing partner interactions with service call tuples. Then the service call tuples list can be analyzed to track the asset 'merging' or 'splitting' activities. 
There are three situations that may incur such activities:

\begin{itemize}
\setlength{\itemsep}{1pt}
\setlength{\parskip}{0pt}
\setlength{\parsep}{0pt}

\item If $X$ sends information containing asset value to $Y$, who aggregates it with its own information (expressed as $Y$ calling itself) and further send it to $Z$. In this situation, we can identify the follow service call tuple sequence:

\begin{equation}
\begin{split}
&<X\stackrel{d}{\longrightarrow}Y, \Delta_X>\\
&<Y\stackrel{d}{\longleftrightarrow}Y, \Delta_X, \Delta_Y>\\
&<Y\stackrel{d}{\longrightarrow}Z, \Delta_Y>
\end{split}
\label{relay}
\end{equation}

\item If $X$ sends information within its request to $Y$ and gets response(s) from $Y$ that includes $X$'s information.  This situation is represented by the follow service call tuple:

\begin{equation}
<X\stackrel{d}{\longleftrightarrow}Y, \Delta_X, \Delta_Y>
\label{includ}
\end{equation}

Extra attentions should be paid in this case, as we can not be sure that the response message includes information from the request message. Whether the output (responses) from a partner integrates the input (request) or not depends on the business logic of this partner's system. An example of this case is when $X$ sends some personal information to $Y$ to calculate the insurance premium. If the response from $Y$ consists in the insurance premium and the person's information, there is an assets derivation, otherwise if $Y$ answers with only the insurance premium, there is no assets derivation. Deciding what information includes 'asset value' and should be protected is closely related to application domain. In any case, we need to know relations between inputs and outputs to conclude if assets derivation exists during a direct interaction. This can be done by analyzing partner's service functional description, e.g. WSDL in a Web Service context. It can also be done at the business process level, by letting the service composition service to adding extra indicators to a WS-BPEL script.
In the modeling level, we use the following notations to define whether the partner response includes information from request or not:
\begin{itemize}
\setlength{\itemsep}{1pt}
\setlength{\parskip}{0pt}
\setlength{\parsep}{0pt}
\item Most of the time, request information (or part of it) is included in the response, therefore we use the default tuple to represent it:
    \begin{equation}
    <Y\stackrel{d}{\longleftrightarrow}Y, \Delta_i, \Delta_o>
    \end{equation}

\item Whereas '$\not\subset$' is used to indicate that no information of the request is included in the response:
    \begin{equation}
    <Y\stackrel{d}{\longleftrightarrow}Y, \Delta_i, \Delta_o, \not\subset>
    \end{equation}
\end{itemize}

\item If $X$ 'fetches' (expressed by '$stackrel{c}{\longrightarrow}$', as there is no asset value in the request) information from $Y$ and aggregates its own information with it:
\begin{equation}
\begin{split}
&<X\stackrel{c}{\longrightarrow}Y>\\
&<X\stackrel{d}{\longleftarrow}Y, \Delta_Y>\\
&<X\stackrel{d}{\longleftrightarrow}X, \Delta_Y, \Delta_X>
\end{split}
\label{fetch}
\end{equation}

\end{itemize}

As an example, we build the list of service call tuples for 'use case 1' (See formula \ref{SCGconstct}, where the tuples in the list are indexed by the steps of business process):

\begin{equation}
\begin{split}
&<\tau 1,D\stackrel{c}{\longrightarrow}B>\\
&<\tau 2,B\stackrel{c}{\longrightarrow}C>\\
&<\tau 3,B\stackrel{d}{\longleftarrow}C,'E'>\\
&<\tau 4,B\stackrel{d}{\longleftrightarrow}B,'E','ME'>\\
&<\tau 5,D\stackrel{d}{\longleftarrow}B,'ME'>
\end{split}
\label{SCGconstct}
\end{equation}

The assets derivation relations between partners are equivalent to data dependency relations between them. Therefore assets derivation trail, which decides the sub-context pattern, can be mined from the list of service call tuples.

\subsection {Sub-context slicing} \label {sc-slic}
Like the information \emph{reachability} questions in SDG, the assets derivation trail can be tracked by scanning the service call tuples list, paying particular attention to asset aggregation. Based on this, providers' policies upon assets can be maintained during assets derivations. This involves, firstly, allocating corelated assets in the same sub-contexts.

We use a data structure 'context development tuple' $<N_C, V_C, P_C, L_A, L_P, S_C>$ to record the information of sub-context development, where:
\begin{itemize}
\setlength{\itemsep}{1pt}
\setlength{\parskip}{0pt}
\setlength{\parsep}{0pt}
\item $N_C$ is the name of the sub-context.
\item $V_C$ is its version.
\item $P_C$ the parent sub-context.
\item $L_A$ a list of all the asset involved in the sub-context.
\item $L_P$ the collection of policies in the sub context.
\item $S_C$ the step of business process.
\end{itemize}

This tuple is built by the sub-context slicing process which scans the SCG (e.g. \emph{service call tuple} list) according two strategies: 'asset-based slicing' and 'request-based slicing'.

\textbf{Asset-based slicing} focuses on capturing the aggregation relation among assets.  Using this method, \textbf{a sub-context is created when the first O-Asset is launched} into the collaborative context by the owner. When a new partner join the context with a new O-Asset, the sub-context consisting of the existing asset is updated, if the new partner's O-Asset is merged with the existing C-Asset. Otherwise (i.e. the new partner's O-Asset is not merged with existing C-Asset), a new sub-context is created. In 'use case 1', the list of sub-context tuples is as follows:

\begin{equation}
\begin{split}
&<R_{CB}, 1, (\phi),(e), (RoP_C), \tau 3>\\
&<R_{CB}, 2, (<R_{CB}.1),(e,m), (RoP_C,RoP_B), \tau 4>\\
&<R_{CB}, 3, (<R_{CB}.2),(e,m), (RoP_C,RoP_B), \tau 5>\\
\end{split}
\label{SClist}
\end{equation}

We can see that in step 3 (represented by $\tau 3$), the first sub-context is created, including the asset $E$ and the $RoP_C$ on it. We name the context after the interaction leading to the creation of it, e.g. $R_{CB}$ ('resource' sending from $C$ to $B$). Its version is '$1$'. It has no parent context ($\phi$). Then in step 4, as new asset $M$ merge with $E$, the sub-context $R_{CB}.1$ is updated to $R_{CB}.2$. And in step 5, it stays unchanged.

\begin{figure}[htbp]
\centering
\includegraphics[width=0.2\textwidth]{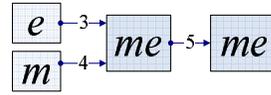}
\caption[Assets derivation in the sample use case]{Assets derivation in the sample use case} \label{fig:graph507}
\end{figure}

This list describes the evolution of the sub-contexts. There is only one sub-context for 'use case 1',  which can be represented with an assets derivation diagram (see figure \ref{fig:graph507}).

Using the 'asset-based slicing' method, policy negotiation and aggregation (including conflicts detection) can not be done until the first asset launched into the context ('step 4' of use case 1). If there is conflict, steps $\tau 2$ and $\tau 3$ are actually wastes of partners' resources and don't need to be proceeded. Therefore the 'asset-based slicing' method is better to be used for 'pre-processing' a business process script (e.g. WS-BPEL documents) before the execution of it. To analyze a collaborative context 'on-the-fly', we need a 'request-based slicing' method.


\textbf{Request-based slicing} creates a sub-context \textbf{when the first request is made}. Then when a new partner joins the business process, its QoP can either be aggregated into existing sub-context, or lead to the creation of a new sub-context. The decision is also straight forward: \textbf{the QoPs of two partners should be aggregated, if they will access the same asset in future steps of the collaboration context}. With this method, we have the following list of sub-context tuples for 'use case 1':
\begin{equation}
\begin{split}
&<Q_{DB}, 1, (\phi), (QoP_D), \tau 1>\\
&<Q_{DB}, 2, (Q_{DB}.1), (QoP_D,QoP_B), \tau 2>
\end{split}
\label{SCQlist}
\end{equation}

This tuple list captures QoP aggregations. When the first request is made by $D$ in step 1, a sub-context is created, including the QoP of D. We name the context after the interaction leading to the creation of it, e.g. $Q_{DB}$ ('query' sending from $D$ to $B$). In step 2, as $B$ is requesting assets from $C$ 'on behalf of' $D$, $QoP_B$ and $QoP_C$ are aggregated. Therefore sub-context $Q_{DB}.1$ is updated to $Q_{DB}.2$

However, deciding who will access the same asset is more tricky than it may firstly look like, especially when partners work asynchronously, (e.g. between that a partner '$X$' receives request from partner '$Y$' and responds $Y$, another partner '$Z$' sends request to $X$). We provide basic protocols for dealing with such cases:
\begin{itemize}
\setlength{\itemsep}{1pt}
\setlength{\parskip}{0pt}
\setlength{\parsep}{0pt}
\item  \textbf{Protocol 1}: \emph{After $X$ receiving request from $Y$, all requests that $X$ sends to other partners are deemed as \textbf{on behalf of} $Y$, until $X$ responds $Y$, or $X$ receives a request from another partner $Z$}. This involves that a request from $Y$ to $X$ establishes a 'on behalf of' relation. Consequently, the $QoP_Y$ should be aggregated into $QoP_X$ for all the requests $X$ sends after receiving request from $Y$, until $X$ gets the result and responds $Y$. The 'on behalf of' relation between $X$ and $Y$ ends when $X$ responds $Y$. It also can, however, be interrupted before $X$ responding $Y$. The following protocols regulate such cases.
\item  \textbf{Protocol 2}: \emph{An '$X$ on behalf of $Y$' relation is interrupted by another '$X$ on behalf of $Z$' relation if $Z$ makes request between $X$ receiving request from $Y$ and $X$ responding $Y$}.

\item  \textbf{Protocol 3}: \emph{An '$X$ on behalf of $Y$' relation interrupted by another request from $Z$ can be resumed after $X$ responds $Z$, if $X$ receives a response from a partner $P$, who was called by $X$ 'on behalf of $Y$'}. This means that the 'on behalf of' relation can be \textbf{nested}. For example, with following request-response sequence (i.e. service call tuple list in formula \ref{smpltocol}), we can say the 'on behalf of' relation between $X$ and $Y$ is restored after '$X$ responds $Z$' (step 5), by the interaction where '$P$ responds $X$', as '$P$' is a partner that $X$ has requested on behalf of $Y$.
    \begin{equation}
    \begin{split}
    &<\tau 1,Y\stackrel{c}{\longrightarrow}X, \Delta_{i1}>\\
    &<\tau 2,X\stackrel{c}{\longrightarrow}P, \Delta_{i2}>\\
    &<\tau 3,Z\stackrel{c}{\longrightarrow}X, \Delta_{i3}>\\
    &<\tau 4,X\stackrel{d}{\longleftrightarrow}Q, \Delta_{i4}, \Delta_{o4}>\\
    &<\tau 5,Z\stackrel{d}{\longleftarrow}X, \Delta_{o3}>\\
    &<\tau 6,X\stackrel{d}{\longleftarrow}P, \Delta_{o2}>\\
    &<\tau 7,Y\stackrel{d}{\longleftarrow}X, \Delta_{o1}>\\
    \end{split}
    \label{smpltocol}
    \end{equation}
\end{itemize}

These are 'basic' protocols because they handle the most simple cases in service composition. When dealing with real-world complex business federations, more information concerning the business process and partner functionalities should be taken into consideration. But the basic reasoning process remains in accordance with those described in these protocols.

In the followings we discuss the employment of 'asset-based' and 'request-based' methods for context slicing. For this, we firstly give an overview of sub-context developments that can occur in a collaborative business process.

\subsection{Context development}\label{cntxtdev}
During each step (partner interaction) of the business process, different types of sub-context development are caused by the partners service calls:

\begin{itemize}
\setlength{\itemsep}{1pt}
\setlength{\parskip}{0pt}
\setlength{\parsep}{0pt}
\item  \textbf{Create}: The creation of a new sub-context is always based on a independent QoP or RoP from a partner. In other words, if the partner provides an asset which is not aggregated with other assets in \emph{current step}, a new sub-context consisting of this asset and the corresponding RoP is created. Analogously, if the partner is calling others '\emph{\textbf{on its own behalf}}' (i.e. not because it is doing so for responding another 'former' requestor) a new sub-context consisting of its QoP should be created.
\item  \textbf{Update}: On the contrary, updating an existing sub-context happens if the partner's asset has data dependency (according to discussions in section \ref{Assagg}) with the assets belonging to an existing sub-context, or if this partners' assets are merged with existing assets. It also happens when the partner is requesting assets on behalf of other 'former' requestor, that is, it's QoP and the QoP of the former requestor should be 'transmitted' to the requested party. Therefore the QoPs are in the same sub-context.
\item  \textbf{Merge}: Merging sub-contexts is a special kind of update operation. It happens when two existing assets in two sub-contexts merge, or when the request sand by a partner is on behalf of several former requestors from different sub-contexts. In the later case, the different sub-contexts are corelated by the asset value in the responding message.
\item  \textbf{Split}: While 'splitting' a sub-context, several new sub-contexts are created. They all 'inherit' the assets and policies of the previous context. Context splitting can be caused by three types of interactions:
    \begin{itemize}
\setlength{\itemsep}{1pt}
\setlength{\parskip}{0pt}
\setlength{\parsep}{0pt}
\item a party sends copies of the same asset to several partners and the copies are developed differently;
\item a party sends copies of the same request to several partners at the same time;
\item the business process has a control structure defining parallel activities.
    \end{itemize}
\item  \textbf{End}: Ending a sub-context occurs when it is \emph{merged}, \emph{split} or when the whole \emph{business process ends}.
\end{itemize}

\subsection {Pre-processing and on-the-fly processing} \label {pro-slic}
In context slicing, \emph{pre-processing} refers to the circumstances where a pre-defined business process (e.g. WS-BPEL script) is analyzed before the execution, to see whether it can be carried out, w.r.t. partners' security profile-request satisfiability. This can be done with the 'asset-based' slicing method, given the policies and attributes of partners.

In 'on-the-fly' processing, partners' RoPs and QoPs must be aggregated as soon as they join the collaboration context, in order to find out conflicts more timingly. This requires using both 'asset-based slicing' and 'request-based slicing'.

For 'use case 1', on-the-fly slicing strategy first builds the QoP tuples (see formula \ref{SCQlist}) from the start of the business process, using 'request-based' slicing. Then from step '$\tau 3$', RoP tuples are built (see formula \ref{SClist}), using 'asset-based' slicing. 

The decided RoP aggregation relations and QoP aggregation relations are used to generate the CSPs of each sub-context. When a new partner joins the collaboration context, it is allocated to a sub-context according to whether it's an asset provider or consumer (or both). Then it's policies are aggregated to the $CSP$ of that sub-context.

Next section gives a more featured illustration of the context slicing method based on a full-fledged collaborative context use case.


\section{Slicing a complex context}\label{GCbp}
This section demonstrates the context slicing method with a more comprehensive use case (see figure \ref{fig:graph508}). First, we briefly introduce  the business process of this use case. Then, the corresponding 'Service Call Graph' and 'service call tuple list' are described, before the 'on-the-fly' slicing process is discussed. Policies and CSPs of this use case are omitted for simplicity (can be found in \cite{ZSu2012a}).

\textbf{Use case 2}: This use case is a collaborative business process for price inquiry. The tourism association 'Eighty days around the World' (E) inquires 'Alice fantasy tourism' (A) for the total price and arrangement of 'Cote d'Azur and the Mediterranean package tour' for 50 persons. 'A' inquires 'Beausoleil tourist office' for fete information. Then 'A' produces the 'coach tour' arrangement information. It further inquires 'Cote d'Azur airline' (C) for travel arrangement (including air transport and accommodation); 'David cruise line' (D) for cruise arrangement; 'Friend-arm' (F) for assurance. 'C' provides the arrangement of 'airline' and inquires three hotels 'Generous' (G), 'Hospitable' (H) and 'Ideal'(I) for room arrangements.

\subsection{Collaborative business process}\label{theBPs}
\begin{figure}[htbp]
\centering
\includegraphics[width=0.5\textwidth]{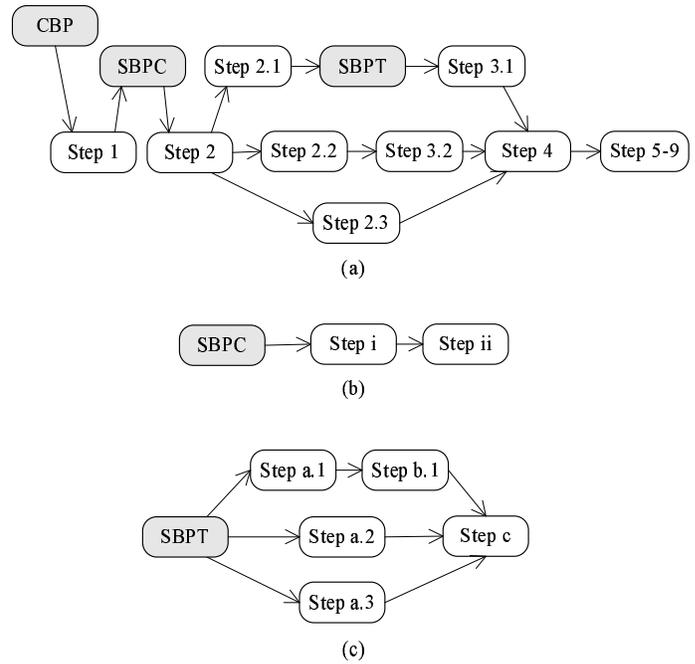}
\caption[Collaborative business process example]{Collaborative business process example: \emph{(a)CBP; (b)SBPC; (c)SBPT.}} \label{fig:graph508}
\end{figure}

The main business process (denoted as 'CBP') comprises 9 steps and two sub-process ('SBPC' and 'SBPT'), as shown in figure \ref{fig:graph508}.

The descriptions of the steps in CBP are as follows:
\begin{itemize}
\setlength{\itemsep}{1pt}
\setlength{\parskip}{0pt}
\setlength{\parsep}{0pt}
\item \textbf{step 1:} 'E' sends 'tourists info' ($o$) to 'A' to query for 'total price and arrangement'.
\item \textbf{step SBPC:} 'A' initiates a sub-process 'SBPC' to to inquiry the 'fete' ($f$) information.
\item \textbf{step 2:} It's composed of three concurrent sub-steps:
\begin{itemize}
\setlength{\itemsep}{1pt}
\setlength{\parskip}{0pt}
\setlength{\parsep}{0pt}
\item \textbf{step 2.1:} 'A' sends $o$ and $f$ to 'C' to ask for 'travel' information ($v$). It is followed (indirectly, there is a sub-process 'SBPT' between them) by two sub-steps:
\begin{itemize}
\setlength{\itemsep}{1pt}
\setlength{\parskip}{0pt}
\setlength{\parsep}{0pt}
\item \textbf{SBPT} 'C' initiates a sub-process 'SBPT' to compose $v$.
\item \textbf{step 3.1:} 'C' sends $v$ to 'A'.
\end{itemize}
\item \textbf{step 2.2:} 'A' sends $o$ to 'D' to query for 'cruise' information ($u$). It is followed by a step:
\begin{itemize}
\setlength{\itemsep}{1pt}
\setlength{\parskip}{0pt}
\setlength{\parsep}{0pt}
\item \textbf{step 3.2:} 'D' sends $u$ to 'A'.
\end{itemize}
\item \textbf{step 2.3:} 'A' provides 'coach tour' information ($k$).
\end{itemize}

\item \textbf{step 4:} 'A' combines $v$, $u$, $k$ and $o$ to 'arrangement' information ($r$).
\item \textbf{step 5:} 'A' sends $r$ to 'F' to query for 'assurance' information ($n$).
\item \textbf{step 6:} 'F' sends $n$ to 'A'.
\item \textbf{step 7:} 'A' combines all the information to 'total price and arrangement' ($t$).
\item \textbf{step 8:} 'A' sends the $t$ to 'E'.
\item \textbf{step 9:} 'E' processes $t$.
\end{itemize}

The steps in \textbf{SBPC} are:
\begin{itemize}
\setlength{\itemsep}{1pt}
\setlength{\parskip}{0pt}
\setlength{\parsep}{0pt}
\item \textbf{step i:} 'A' queries 'B' for $f$.
\item \textbf{step ii:} 'B' sends $f$ to 'A'.
\end{itemize}

The steps in \textbf{SBPT} are:
\begin{itemize}
\setlength{\itemsep}{1pt}
\setlength{\parskip}{0pt}
\setlength{\parsep}{0pt}
\item \textbf{step a:} It consists in three concurrent steps initiated by 'C':
\begin{itemize}
\setlength{\itemsep}{1pt}
\setlength{\parskip}{0pt}
\setlength{\parsep}{0pt}
\item \textbf{step a.1:} 'C' sends $o$ to 'G' to ask for 'room' information ($m$). It is followed by:
\begin{itemize}
\setlength{\itemsep}{1pt}
\setlength{\parskip}{0pt}
\setlength{\parsep}{0pt}
\item \textbf{step b.1:} 'G' sends $m_G$ information to 'C'.
\end{itemize}
\item \textbf{step a.2:} policy negotiation shows that the \(QoP_{H}\) does not satisfy \(RoP_{E}\); 'C' ceases calling the service of 'H'.
\item \textbf{step a.3:} policy negotiation shows that the \(QoP_{E}\) does not satisfy \(RoP_{I}\); 'I' refuses to work with 'C', because 'C' is requesting 'on behalf of' E.
\end{itemize}
\item \textbf{step c:} 'C' combines 'airline' (l) information with '$m_G$' to produce (v).
\end{itemize}

In the following sections, we build the Service Call Diagram (SCG) and discuss the sub-context slicing process.

\subsection{Service Call Graph}\label{theSCG}
\begin{figure}[htbp]
\centering
\includegraphics[width=0.5\textwidth]{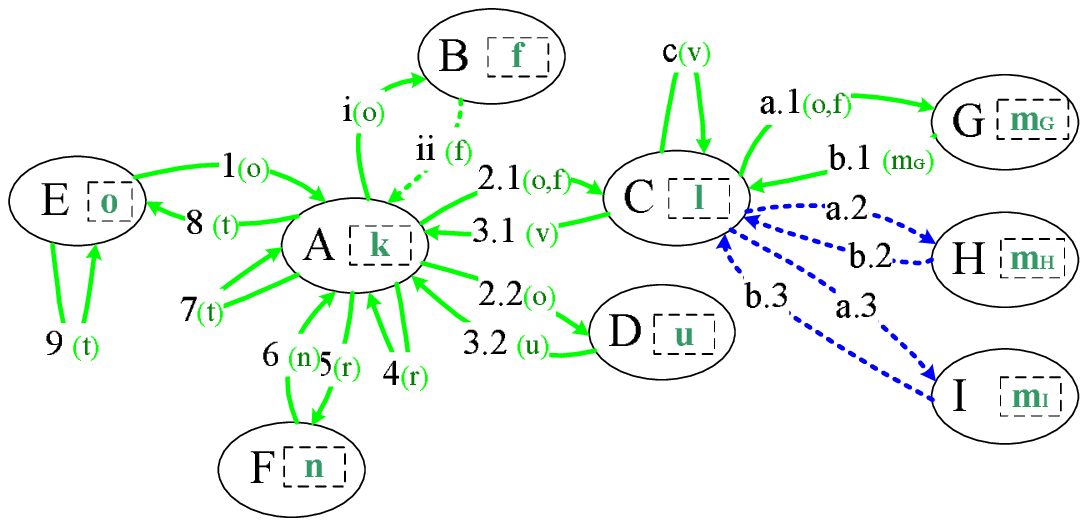}
\caption[SCG of the collaborative business process]{SCG of the collaborative business process. \emph{Meanings of elements are:}\\
\begin{minipage}{0.5\textwidth}\emph{
\begin{itemize}
\setlength{\itemsep}{1pt}
\setlength{\parskip}{0pt}
\setlength{\parsep}{0pt}
\item Green lines represent data-exchange messages which carries assets, where
\begin{itemize}
\setlength{\itemsep}{1pt}
\setlength{\parskip}{0pt}
\setlength{\parsep}{0pt}
\item Solid green lines (1, i, 2.1, a.1 b.1, c, 3.1, 2.2, 3.2, 4, 5, 6, 7, 8, 9) represent the output messages that have used information from the input messages.
\item Dashed green lines (ii) represent the output messages that don't include information from the input.
\end{itemize}
\item Solid Blue lines (a.2, b.2, a.3, b.3) represent control messages which doesn't carry assets.
\item Dashed blue lines represent the interaction that failed, due to negative result of policy negotiation.
\item Each green line is attached with the asset it carries.
\end{itemize}}
\end{minipage}
} \label{fig:graph509}
\end{figure}

The service call diagram (figure \ref{fig:graph509}) shows the partner interaction during these steps.


The list of service call tuple is as in formula (\ref{CSCGconstct}):
\begin{equation}
\begin{split}
<&\tau 1,E\stackrel{d}{\longrightarrow}A,(o)>\\
<&\tau i+ii,A\stackrel{d}{\longleftrightarrow}B,(o),(f), \not\subset>\\
<&\tau 2.1,A\stackrel{d}{\longrightarrow}C,(o)>\\
<&\tau a.1+b.1,C\stackrel{d}{\longleftrightarrow}G,(o),(m_G)>\\
<&\tau a.2+b.2,C\stackrel{f}{\longleftrightarrow}H,\not\subset>\\
<&\tau a.3+b.3,C\stackrel{f}{\longleftrightarrow}I,\not\subset>\\
<&\tau c,C\stackrel{d}{\longleftrightarrow}C,(m_G,l),(v)>\\
<&\tau 3.1,A\stackrel{d}{\longleftarrow}C,(v)>\\
<&\tau 2.2+3.2,A\stackrel{d}{\longleftrightarrow}D,(o),(u)>\\
<&\tau 4,A\stackrel{d}{\longleftrightarrow}A,(v,u,k),(r)>\\
<&\tau 5+6,A\stackrel{d}{\longleftrightarrow}F,(r),(n)>\\
<&\tau 7,A\stackrel{d}{\longleftrightarrow}A,(r,n),(t)>\\
<&\tau 8,E\stackrel{d}{\longleftarrow}A,(t)>\\
<&\tau 9,E\stackrel{d}{\longleftrightarrow}E,(t)>\\
\end{split}
\label{CSCGconstct}
\end{equation}

\subsection{On-the-fly slicing}\label{slcing}
The business process starts with E's request, which leads to the creation of two sub-context. First, a QoP aggregation starts from the $QoP_E$. Second, a RoP aggregation starts from $RoP_E$, as the request carries the 'tourist information' (o).

\subsubsection{QoP aggregation}\label{QoPag}
Firstly, we track the QoP aggregation process. By step 1, $QoP_E$ leads to the creation of a sub-context:

\begin{equation}
<Q_{EA}, 1, (\phi), (QoP_E), \tau 1>
\label{QoPofE}
\end{equation}

\begin{figure}[htbp]
\centering
\includegraphics[width=0.5\textwidth]{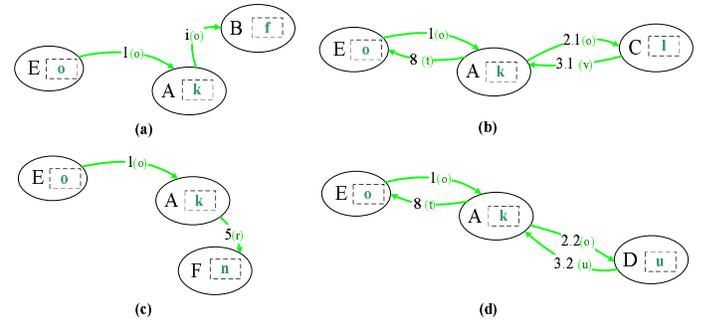}
\caption[QoP aggregation]{QoP aggregation.\emph{ (a) E, A, B (b) E, A, C (c) E, A, F (d) E, A, D}} \label{fig:graph510}
\end{figure}

This sub-context splits as 'A' calls 'B', 'C', 'D' and 'F' separately. Thus we have 4 parallel sub-contexts created (see figure \ref{fig:graph510}).

They are presented by the following sub-context development tuples:
\begin{equation}
\begin{split}
&<Q_{AB}, 1, (Q_{EA}.1), (QoP_E,QoP_A), \tau i>\\
&<Q_{AC}, 1, (Q_{EA}.1), (QoP_E,QoP_A), \tau 2.1>\\
&<Q_{AD}, 1, (Q_{EA}.1), (QoP_E,QoP_A), \tau 2.2>\\
&<Q_{AF}, 1, (Q_{EA}.1), (QoP_E,QoP_A), \tau 5>\\
\end{split}
\label{QoPEA}
\end{equation}

\begin{figure}[htbp]
\centering
\includegraphics[width=0.45\textwidth]{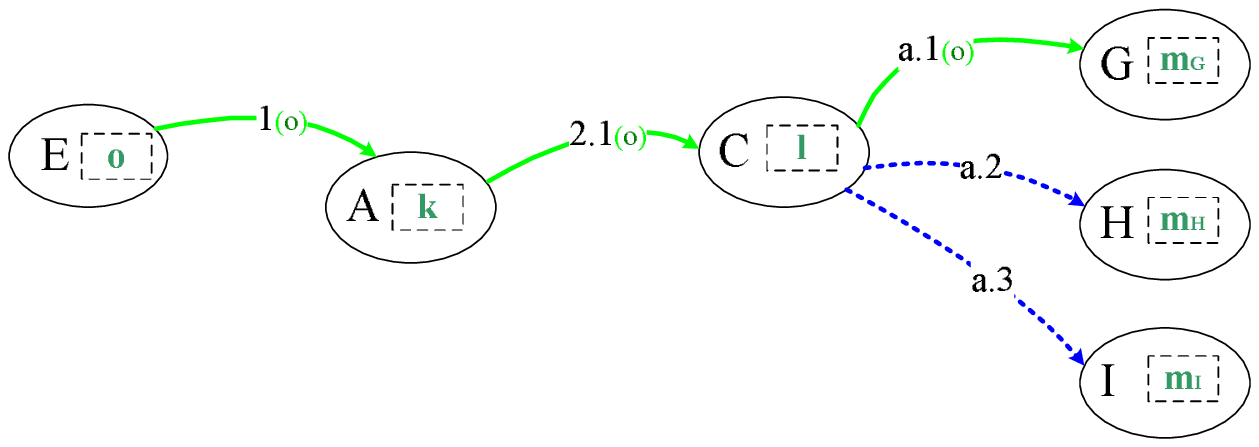}
\caption[Details of QoP aggregation along 'E, A, C']{Details of QoP aggregation along 'E, A, C'} \label{fig:graph511}
\end{figure}

The sub-context $QoP_C.1$ further splits into 3 three new sub-contexts, as 'C' calls 'G', 'H' and 'I' for information (see figure \ref{fig:graph511}).

\begin{equation}
\begin{split}
&<Q_{CG}, 1, (Q_{AC}.1), (QoP_E,QoP_A,QoP_C), \tau a.1>\\
&<Q_{CH}, 1, (Q_{AC}.1), (QoP_E,QoP_A,QoP_C), \tau a.2>\\
&<Q_{CI}, 1, (Q_{AC}.1), (QoP_E,QoP_A,QoP_C), \tau a.3>\\
\end{split}
\label{QoPCG}
\end{equation}

Set that $QoP_E$ doesn't satisfy $RoP_H$ and $RoP_I$, This list of sub-context development tuples help us to find out this right after 'H' or 'I' is called. Otherwise, if using only RoP aggregation, we will find out that $QoP_E$ doesn't meet $RoP_I$ only at 'step 8', when the artifacts consisting asset from 'H' or 'I' are send to 'E'.

In short, the QoP aggregation 'transmits' the up-stream requesters' QoPs along the business process, so policies which do not match can be found in time. However, QoP aggregation isn't sufficient. As E's request carries $o$, any downstream partners receiving the request should fulfill $RoP_E$.


\subsubsection{RoP aggregation}\label{RoPag}
We start to track RoP aggregation process from step 1, when 'E' send its request carrying $o$:
\begin{equation}
<R_{EA}, 1, (\phi), (o), (RoP_E), \tau 1>
\label{RoPofE}
\end{equation}

By 'step ii', as B's response $f$ doesn't contain $o$, there is no asset aggregation between 'B' and others in this step. Therefore, a new sub-context is created.

\begin{equation}
<R_{BA}, 1, (R_{EA}.1), (f), (RoP_B), \tau ii>
\label{RoPofB}
\end{equation}

However, these two sub-contexts should merge, as 'A' aggregates $f$ and $o$ before step 2.1 and 2.2.
\begin{equation}
\begin{split}
<&R_{EA}, 2, (R_{EA}.1,R_{BA}.1), (o,f), (RoP_E,RoP_B), \tau ii>
\end{split}
\label{RoPofEB}
\end{equation}

This sub-contexts are succeeded by two different sub-contexts, as 'A' calls 'C' and 'D' in parallel.

\begin{itemize}
\setlength{\itemsep}{1pt}
\setlength{\parskip}{0pt}
\setlength{\parsep}{0pt}
\item
After 'A' calling 'C', 'C' in turn calls 'G' and get $m_G$. Then, 'C' aggregate $m_G$ with $l$ to get $v$. Thus we have the sub-context tuples:
\begin{equation}
\begin{split}
<R_{GC}, 1, &(R_{EA}.2), (m_G), \\
&(RoP_E,RoP_B,RoP_G), \tau b.1>\\
<R_{GC}, 2, &(R_{GC}.1), (m_G,l), \\
&(RoP_E,RoP_B,RoP_G,RoP_C), \tau c>\\
\end{split}
\label{RoPofEBCG}
\end{equation}

\item
With 'A' calling 'D' and getting $u$, we have:
\begin{equation}
\begin{split}
<R_{DA}, 1, &(R_{EA}.2), (u), \\
&(RoP_E,RoP_B,RoP_D), \tau 3.2>\\
\end{split}
\label{RoPofEBD}
\end{equation}
\end{itemize}

 These two sub-contexts are merged when 'A' merges $u$, $v$ and $k$ into $r$.
\begin{equation}
\begin{split}
<&R_{DA}, 2, (R_{GC}.2,R_{DA}.1), (u,v,k),\\
&(RoP_E,RoP_B,RoP_G,RoP_C,RoP_D, RoP_A), \tau 4>
\end{split}
\label{RoPofAag}
\end{equation}

Now 'A' calls 'F' with asset $r$ and the aggregated RoP of 'E', 'B', 'G', 'C', 'D' and 'A'. Set that $QoP_F$ fulfills the aggregated RoP, 'F' will join the context, providing asset $n$. Therefore the sub-context $R_{DA}.2$ get updated.

\begin{equation}
\begin{split}
<R_{DA}, 3, &(R_{DA}.2), (n),(RoP_E,RoP_B,RoP_G,\\&RoP_C,RoP_D, RoP_A, RoP_F),\tau 6>\\
\end{split}
\label{RoPoffag}
\end{equation}

Then 'A' merges $n$ with $r$ into $t$.

\begin{equation}
\begin{split}
<R_{DA}, 4, &(R_{DA}.3), (n,r),(RoP_E,RoP_B,RoP_G,\\&RoP_C,RoP_D, RoP_A, RoP_F), \tau 7>
\end{split}
\label{RoPofall}
\end{equation}

In step 8 the aggregated RoP in sub-context $R_{DA}.4$ (formula \ref{RoPofall}) is used to exam $QoP_E$.

The on-the-fly strategy analyze the business process to produce QoP and RoP aggregations. QoP aggregation allows transmitting former requesters' security attributes down-stream, to match RoPs in time. RoP aggregation propagates former providers' security requirements down-stream, to ensure no leakage of the protected information to unauthorized consumer.

\section{Implementation}\label{Implmt}
\subsection{'Context manager' component}\label{CM}
\begin{figure}[htbp]
\centering
\includegraphics[width=0.4\textwidth]{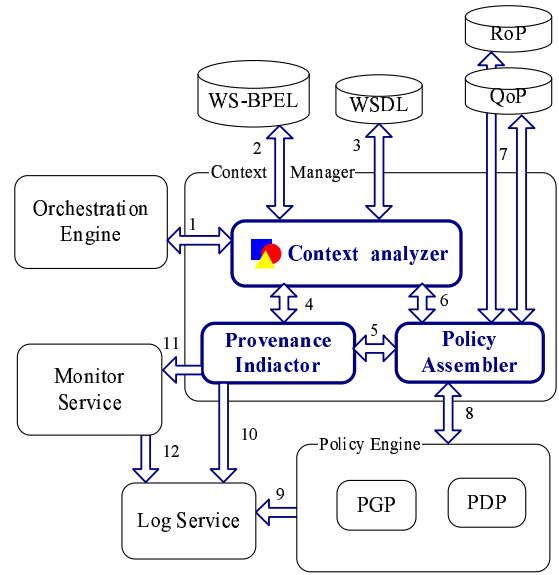}
\caption[Components of context manager]{Components of context manager} \label{fig:graph603}
\end{figure}

Implementation involves a 'context manager' service (see figure \ref {fig:graph603}), which tracks the assets derivation to decide policy aggregation relations and RoP-QoP negotiation relations among partners. Such information is sent to the policy engine we developed in former work \cite{ZSuFB2012a}, which includes a Policy Decision Point (standard XACML PDP, as our policy model can be implemented with XACML) for negotiation and a Policy Gathering Point for policy aggregation.

Our context manager cooperates with the business engine to get the business process description encoded with WS-BPEL, for 'pre-processing' it before the 'orchestration engine' starts business session. Major steps during the processing are: 
\begin{itemize}
\setlength{\itemsep}{1pt}
\setlength{\parskip}{0pt}
\setlength{\parsep}{0pt}
\item step 1: The 'context analyzer' loads a business process defined with WS-BPEL and fetches the WSDL files of the business partners defined in the WS-BPEL.

\item step 2: It uses 'asset-based' slicing method to proceed sub-contexts partition and allocates assets to them, according to the assets merging/ inheriting relations. The 'provenance indicator' data records the sub-context development information.

\item step 3:  For each 'sub context', the 'policies assembler' fetches the RoP and QoP policies of all partners, indicated by their WSDLs. It parses the QoPs to complete the requests. Then it sends requests and RoPs to the 'Negotiation and Aggregation' engine ('PDP' for negotiation, 'PGP' for aggregation.

\item step 4: If the negotiation and aggregation results for all the 'sub contexts' are positive, the 'context manager' will call the 'orchestration engine' to start the business process.
\end{itemize}

%
%

The analyzing algorithm is organized depending on the '$<activities>$' elements in WS-BPEL  scripts, due to their impacts on sub-context development.
Basically, they can be differentiated into 4 categories:
\begin{itemize}
\setlength{\itemsep}{1pt}
\setlength{\parskip}{0pt}
\setlength{\parsep}{0pt}
\item \textbf{category 1}:  The \emph{activities} that lead to information exchanges between partners, including $<receive>$, $<reply>$, $<invoke>$, $<assign>$ and $<exit>$. This kind of \emph{activities} can only result in the create/update/merge/end of context.
\item \textbf{category 2}: The 'control' \emph{activities} that lead to context split, including: $<sequence>$; $<flow>$ with '$parallel=yes$' factor; $<forEach>$ with '$parallel=yes$' factor; $<if>$ with \textbf{different} \emph{activities} in its branches.
\item \textbf{category 3}: The 'control' \emph{activities} that do not lead to context split, including $<pick>$, $<scope>$, $<while>$ and $<repeatUntil>$. But their children \emph{activities} should be examined.
\item \textbf{category 4}: The \emph{activities} that are irrelevant to context slicing, including  $<throw>$, $<wait>$, $<empty>$, $<compensate>$, $<compensateScope>$, $<rethrow>$, $<validate>$, $<extensionActivity>$.
\end{itemize}

The context slicing process is carried out by two methods. Method 'coordinator'  (see algorithm \ref{crdtr}) locates the process starting point, i.e. a '$<receive>$', '$<pick>$' or '$<onEvent>$' activity that has a '$createInstance$' factor, and creates the first context. Then it gets all the following \emph{activities} and sends them to the 'analyze' method for tracking asset aggregations. 

\begin{algorithm}[htb]
\caption{method 'coordinator'}
\label{crdtr}
\begin{algorithmic}[1]
\REQUIRE ~~\\
    The business process defined with WS-BPEL\\
\ENSURE ~~\\
    An 'assembler' Object, which comprises the context slicing information
\STATE Locate the starting $activity$ of business process;
\STATE Create the first '$asset$' object according to the starting activity;
\STATE Create the first '$context$' object with the first '$asset$' object, the '$variable$' and '$step$' information;
\STATE Create the list of '$context$', which consists only the current $context$;
\STATE Set the value of '$slicer$' to $update$;
\FOR {Each $activity$ that follows starting activity}
     \STATE Call method 'analyze' with the $activity$, the '$slicer$' and the list of $context$;
     \STATE Get the updated list of $context$;
\ENDFOR
\STATE create 'assembler' object with the list of $context$;
\STATE Enrich the $context$ in the list with 'RoPs' and 'QoPs';
\STATE Output the configuration file with information in the list of $context$;
\end{algorithmic}
\end{algorithm}

Method 'analyze'  (see algorithm \ref{anlz}) deals with each activity according to its impact on context development. For activities of category 1, it calls a method 'develop' with '$slicer=update$'
For activities of category '2', it splits ($slicer=split$) the context and examines the children activities.
For activities of category '3', it examines the children activities.

\begin{algorithm}[htb]
\caption{method 'analyze'}
\label{anlz}
\begin{algorithmic}[1]
\REQUIRE ~~\\
    A list of the $Indicator$ object, $slicer$ and current $activity$\\
\ENSURE ~~\\
    An updated list of the $Indicator$ object, $slicer$
\IF {$activity$ in 'category 1'}
     \STATE Call method '\emph{\textbf{develop}}' with this $activity$, the '$context$' list and $slicer$;
\ELSE
    \IF {($activity$ in 'category 2')}
        \FOR {Each child $activity$}
            \STATE Set $slicer='split'$;
            \STATE Call method 'analyze' with the $activity$, the list of $context$ and $slicer$;
        \ENDFOR
    \ENDIF
\ELSE
    \IF {($activity$ in 'category 3')}
        \FOR {Each child $activity$}
            \STATE Call method 'analyze' with the $activity$, the '$context$' list and $slicer$;
        \ENDFOR
    \ENDIF
\ENDIF
\STATE Output the $context$ list;
\end{algorithmic}
\end{algorithm}

The method 'develop' updates the $context$ list according to the '$slicer$' factor. It creates a new '$context$' if $slicer='split'$, updates the version of an existing $context$ if $slicer='update'$.

\subsection{Performance testing}\label{profiling}

This section analyzes the performance of the 'Context Slicing' component, based on experiment with the five 'Sample Processes' taken from the WS-BPEL2.0 specification \cite{WSBPEL2007} ('initial sample' and other 4 samples in section '15 Examples').

\subsubsection{Testing with 'TPTP'}\label{testcmTPTP}
Tables (\ref{tab:table3}) and (\ref{tab:table4}) show the TPTP profiling results of 'execution time' analysis and 'memory consumption' analysis separately. They are both based on running the 'initial sample'.

\begin{table}[htbp]
\centering
\begin{tabular}{|l|l|l|l|}\hline
{\textbf{\parbox[t]{0.5in}{Package}}}
&{\textbf{\parbox[t]{0.55in}{Base Time (ms)}}}
&{\textbf{\parbox[t]{0.55in}{Average Time (ms)}}}
&{\textbf{\parbox[t]{0.55in}{Cumulative Time (ms)}}}
\\\hline
Engine  &1.93 &1.93 &100 \\\hline
I/O  &76.31 &38.15 &76.31 \\\hline
Analyzer &21.13 &0.17 &98.07 \\\hline
Context &0.44 &0.01 &0.44 \\\hline
PartnerLink &0.09 &0.01 &0.09 \\\hline
Variable &0.10 &0.01 &0.10 \\\hline
\end{tabular}
\caption[Time percentage of components' execution]{Time percentage of components' execution} \label{tab:table3}
\end{table}

As shown in table (\ref{tab:table3}), the I/O operation (the method 'getJdomDoc') takes $76\%$ of the total time. The analysis process itself only takes $24\%$ the total time. As the file size of a BPEL document is usually very small, the I/O time doesn't change much with different BPEL files. 
Therefore, we can conclude that the algorithm scales well with complex (and long) BPEL processes.


\begin{table}[htbp]
\centering
\begin{tabular}{|l|l|l|l|l|}\hline
{\textbf{\parbox[t]{0.5in}{Class name}}}
&{\textbf{\parbox[t]{0.4in}{Live Instance}}}
&{\textbf{\parbox[t]{0.55in}{Active Size (byte)}}}
&{\textbf{\parbox[t]{0.43in}{Total Instance}}}
&{\textbf{\parbox[t]{0.54in}{Total Size (byte)}}}
\\\hline
Analyzer &1 &224 &1 &224 \\\hline
Context &16 &896 &16 &896 \\\hline
PartnerLink &4 &128 &4 &128 \\\hline
Variable &5 &160 &5 &160 \\\hline
\end{tabular}
\caption[Memory consumption of components' execution]{Memory consumption of components' execution} \label{tab:table4}
\end{table}

As shown in table (\ref{tab:table4}), the memory consumption of the context slicing method is insignificant. The total size of memory consumption by all the instances is 1440 bytes. Therefore, its impact is trivial when deployed with an orchestration service.

\subsubsection{Deployment testing}\label{testcmDply}
\begin{figure}[htbp]
\centering
\includegraphics[width=0.5\textwidth]{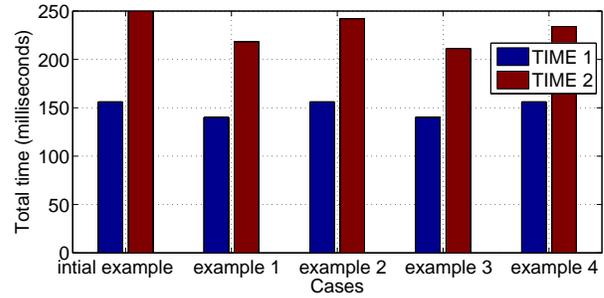}
\caption[Performance of deployment on different environments]{Performance of deployment on different environments} \label{fig:graph622}
\end{figure}

Figure (\ref{fig:graph622}) shows the performance testing results, in terms of execution time for processing the 5 sample WS-BPEL processes. The two environments are:
\begin{itemize}
\setlength{\itemsep}{1pt}
\setlength{\parskip}{0pt}
\setlength{\parsep}{0pt}
\item
'Time1': 'Intel T7250 (Dual Core 2.0 GHz)' CPU and '3.49G' RAM.
\item
'Time2': 'Intel T2330 (Dual Core 1.6 GHz)' CPU and '2.00G' RAM.
\end{itemize}

We can see from figure (\ref{fig:graph622}) that, first, although the business processes are different in length and complexity (see table \ref{tab:table2} for comparison), the processing time for them only varies slightly. Second, the performance changes on different environments are evident.

Detail information of the 5 processes is shown in table (\ref{tab:table2}).

Meanings of column names are:
\begin{itemize}
\setlength{\itemsep}{1pt}
\setlength{\parskip}{0pt}
\setlength{\parsep}{0pt}
\item The column 'partnerLink' describes the number of 'partnerLink' element in each BPEL file,
\item 'variable' the number of 'variable' element,
\item 'basic-activity' the total number of activities that incur partner interaction, i.e. 'receive', 'invoke', 'copy' 'reply' and 'assign'.
\end{itemize}

\begin{table}[htbp]
\centering
\begin{tabular}{|l|l|l|l|l|l|}\hline
{\textbf{\parbox[t]{0.35in}{Examples}}}
&{\textbf{\parbox[t]{0.35in}{Time1\\(ms)}}}
&{\textbf{\parbox[t]{0.35in}{Time2\\(ms)}}}
&{\textbf{\parbox[t]{0.35in}{partn-\\erLink}}}
&{\textbf{\parbox[t]{0.35in}{Varia-\\bles}}}
&{\textbf{\parbox[t]{0.35in}{Basic\\activity}}}
\\\hline
Initial  &156 &250 &4 &5 &10\\\hline
Example1 &140 &218 &1 &3 &9 \\\hline
Example2 &156 &242 &5 &7 &18 \\\hline
Example3 &140 &211 &3 &3 &5 \\\hline
Example4 &156 &234 &3 &6 &14 \\\hline
\end{tabular}
\caption[Detail information of the sample BPEL processes]{Detail information of the sample BPEL processes} \label{tab:table2}
\end{table}

%

We can see from table (\ref{tab:table2}) that the 'Initial example' and 'Example2' are more complex BPEL processes. Correspondingly, their processing time (shown in figure \ref{fig:graph622}) are longer.

\section{Conclusion and future work}\label{Finalcnln}
Originator Control (ORCON) requires downstream consumers to fulfill the originator's requirements, therefore providing full lifecycle protection for assets (i.e. service and information) throughout the whole collaborative context. It ensures that the assets transmission, consumption and propagation are always in the scope defined with providers policies.

Based on analyzing the characteristics of collaborative business process, this paper describes a originator usage control method fitting to inter-enterprise level business federation. We propose a 'Service Call Graph (SCG)' and a corresponding data structure 'service call tuple', based on extending the System Dependency Graph (SDG), to capture 'asset aggregation (and derivation)' in a collaborative business process. A 'context slicing' operation can be made based on the 'SCG', to categorize partners that have direct and indirect assets exchange relations to the same 'sub-contexts'. Security policy negotiation and aggregation done in the scope of each sub-context can ensure originator control principle. A detail discussion is given on the rational of our method, facilitated by two sample use cases. Basically, 'data dependency' between partners incurs assets (and RoP policies) aggregation, whereas 'control dependency' between partners leads to the 'on behalf of' relation and QoP aggregation.
According to data dependency, 'asset-based' slicing is sufficient for 'pre-processing' a business process script (e,g. WS-BPEL script). But for 'on-the-fly processing' a business federation (e.g. dynamic service composition), both 'request-based' (due to control dependency) and 'asset-based' slicing should be used.
Implementation works consolidate a 'context management' service, paying particular attention to the analysis of BPEL-defined business processes. Impacts on the analysis algorithms from different types of '$activity$' defined in WS-BPEL 2.0 specification are discussed. Performance testing results, based on the 'sample business processes' that come with the 'WS-BPEL 2.0' specification, are presented. Future work involves, firstly, the enrichment of context manager functionality with 'on-the-fly' analysis capability, by managing dynamic service composition. This can be done based on 'PEtALS ESB' system, by extending its service management functionality. Following work includes the development of components for enforcing originators usage control policies.

\ifCLASSOPTIONcaptionsoff
  \newpage
\fi

\bibliographystyle{plain}
\bibliography{p2}

\begin{thebibliography}{10}

\bibitem{Bandhakavi-2006-p51-58}
Sruthi Bandhakavi, Charles~C. Zhang, and Marianne Winslett.
\newblock Super-sticky and declassifiable release policies for flexible
  information dissemination control.
\newblock In {\em Proceedings of the 5th ACM workshop on Privacy in electronic
  society}, WPES '06, pages 51--58, New York, NY, USA, 2006. ACM.

\bibitem{LGF2010rohtua}
Laurent Bussard, Gregory Neven, and Franz-Stefan Preiss.
\newblock Downstream usage control.
\newblock In {\em Proceedings of the 11th IEEE International Symposium on
  Policies for Distributed Systems and Networks}, POLICY '10, pages 22--29,
  Washington, DC, USA, 2010. IEEE Computer Society.

\bibitem{Chadwick-2008-p1-6}
David~W. Chadwick and Stijn~F. Lievens.
\newblock Enforcing "sticky" security policies throughout a distributed
  application.
\newblock In {\em MidSec '08: Proceedings of the 2008 workshop on Middleware
  security}, pages 1--6, New York, NY, USA, 2008. ACM.

\bibitem{Daniele2009}
Catteddu Daniele and Hogben Giles.
\newblock {Cloud Computing}: Benefits, risks and recommendations for
  information security.
\newblock Technical report, European Network and Information Security Agency
  (ENISA), November 2009.

\bibitem{ORCON0}
{Director of Central Intelligence}.
\newblock Control of dissemination of intelligence information, directive no.
  1/7., may 4, 1981., 5 1981.

\bibitem{grammatec}
GrammaTech.
\newblock Dependence graphs and program slicing-codesurfer technology overview.
\newblock Technical report, GrammaTech, Inc.

\bibitem{DBLPGuDDXM08}
Liang Gu, Xuhua Ding, Robert~Huijie Deng, Bing Xie, and Hong Mei.
\newblock Remote attestation on program execution.
\newblock In {\em STC}, pages 11--20, 2008.

\bibitem{Hilty-2008-p531-546}
M.~Hilty, A.~Pretschner, D.~Basin, C.~Schaefer, and T.~Walter.
\newblock A policy language for distributed usage control.
\newblock In Joachim Biskup and Javier Lopez, editors, {\em Computer Security
  – ESORICS 2007}, volume 4734 of {\em Lecture Notes in Computer Science},
  pages 531--546. Springer Berlin / Heidelberg, 2007.

\bibitem{Hilty2006}
Manuel Hilty, Alexander Pretschner, Christian Schaefer, and Thomas Walter.
\newblock Enforcement for usage control- an overview of control mechanisms.
\newblock {\em DoCoMo Euro-Labs Publication}, 2006.

\bibitem{Jay2008}
Heiser Jay and Nicolett Mark.
\newblock Assessing the security risks of {Cloud Computing}.
\newblock Technical Report G00157782, Gartner Inc., June 2008.

\bibitem{kagal10Jul}
Lalana Kagal and Hal Abelson.
\newblock Access control is an inadequate framework for privacy protection.
\newblock In {\em W3C Privacy Workshop}. W3C, July 2010.

\bibitem{Karat:2009:PFS:1850636.1850640}
J.~Karat, C.-M. Karat, E.~Bertino, N.~Li, Q.~Ni, C.~Brodie, J.~Lobo, S.~B.
  Calo, L.~F. Cranor, P.~Kumaraguru, and R.~W. Reeder.
\newblock Policy framework for security and privacy management.
\newblock {\em IBM J. Res. Dev.}, 53:242--255, March 2009.

\bibitem{Linda2010}
Ban~B. Linda, Cocchiara Richard, Lovejoy Kristin, Telford Ric, and Ernest Mark.
\newblock The evolving role of {IT} managers and {CIO}s--findings from the 2010
  {IBM} global {IT} risk study.
\newblock Technical report, {IBM}, 2010.

\bibitem{McCollum1990}
Catherine~Jensen McCollum, Judith~R. Messing, and LouAnna Notargiacomo.
\newblock Beyond the pale of mac and dac--defining new forms of access control.
\newblock {\em Security and Privacy, IEEE Symposium on}, 0:190, 1990.

\bibitem{Mont-2003-p377-377}
Marco~Casassa Mont, Siani Pearson, and Pete Bramhall.
\newblock Towards accountable management of identity and privacy: Sticky
  policies and enforceable tracing services.
\newblock In {\em DEXA '03: Proceedings of the 14th International Workshop on
  Database and Expert Systems Applications}, page 377, Washington, DC, USA,
  2003. IEEE Computer Society.

\bibitem{Ni2011}
Qun Ni and Elisa Bertino.
\newblock {xfACL}: an extensible functional language for access control.
\newblock In {\em Proceedings of the 16th ACM symposium on Access control
  models and technologies}, SACMAT '11, pages 61--72, New York, NY, USA, 2011.
  ACM.

\bibitem{WSBPEL2007}
{OASIS}.
\newblock {Web services Business Process Execution Language (WS-BPEL)}.
\newblock http://docs.oasis-open.org/wsbpel/2.0/wsbpel-v2.0.html, April 2007.

\bibitem{Park:2002:OCU:863632.883494}
J.~Park and R.~Sandhu.
\newblock Originator control in usage control.
\newblock In {\em Proceedings of the 3rd International Workshop on Policies for
  Distributed Systems and Networks (POLICY'02)}, POLICY '02, pages 60--,
  Washington, DC, USA, 2002. IEEE Computer Society.

\bibitem{Park2004}
Jaehong Park and Ravi Sandhu.
\newblock The {UCON$_{\mbox{ABC}}$} usage control model.
\newblock {\em ACM Trans. Inf. Syst. Secur.}, 7:128--174, February 2004.

\bibitem{Pretschner2006}
Alexander Pretschner, Manuel Hilty, and David Basin.
\newblock Distributed usage control.
\newblock {\em Commun. ACM}, 49(9):39--44, 2006.

\bibitem{Sandhu1992}
Ravi~S. Sandhu.
\newblock The typed access matrix model.
\newblock In {\em Proceedings of the 1992 IEEE Symposium on Security and
  Privacy}, SP '92, pages 122--, Washington, DC, USA, 1992. IEEE Computer
  Society.

\bibitem{ZSu2012a}
Ziyi Su.
\newblock {\em Applying Digital Rights Management to Corporate Information
  Systems}.
\newblock PhD thesis, INSA Lyon, 20 Avenue Albert Einstein, Villeurbanne,
  France, Mars 2012.

\bibitem{ZSuFB2011c}
Ziyi Su and F\'ed\'erique Biennier.
\newblock Full lifecycle resource protection in composite web service with
  xacml.
\newblock {\em ICIC Express Letters}, 2(5):1045--1050, 2011.

\bibitem{ZsuFB2012b}
Ziyi Su and F\'ed\'erique Biennier.
\newblock An architecture for implementing ’collaborative usage control’
  policy -toward end-to-end security management in collaborative computing.
\newblock In {\em 14th International Conference on Enterprise Information
  System}, ICEIS2012, (submitted).

\bibitem{ZSuFB2012a}
Ziyi Su and F\'ed\'erique Biennier.
\newblock Toward collaborative usage control.
\newblock {\em ACM Trans. Inf. Syst. Secur.}, (submitted).

\bibitem{zsfbAPMS11}
Ziyi Su and Fr\'ed\'erique Biennier.
\newblock Toward comprehensive security policy governance in collaborative
  enterprise.
\newblock In {\em Proceedings of the 7th International Conference on Advances
  in Production Management Systems}, APMS 2011. IFIP WG5.7, 2011.

\bibitem{Biennier2010}
Ziyi Su and Fr\'{e}d\'{e}rique Biennier.
\newblock End-to-end security policy description and management for
  collaborative system.
\newblock {\em Journal of Information Assurance and Security}, (to appear).

\bibitem{Cranor2002}
W3C.
\newblock The platform for privacy preferences 1.0 ({P3P1.0}) specification.
\newblock http://www.w3.org/TR/P3P/, 2002.

\bibitem{Zhang-2006-p180-189}
Xinwen Zhang, Masayuki Nakae, Michael~J. Covington, and Ravi Sandhu.
\newblock A usage-based authorization framework for collaborative computing
  systems.
\newblock In {\em Proceedings of the 11th ACM symposium on Access control
  models and technologies}, SACMAT '06, pages 180--189, New York, NY, USA,
  2006. ACM.

\bibitem{Zhang2008a}
Xinwen Zhang, Masayuki Nakae, Michael~J. Covington, and Ravi Sandhu.
\newblock Toward a usage-based security framework for collaborative computing
  systems.
\newblock {\em ACM Trans. Inf. Syst. Secur.}, 11:3:1--3:36, February 2008.

\bibitem{Zhang-2005-p351-387}
Xinwen Zhang, Francesco Parisi-Presicce, Ravi Sandhu, and Jaehong Park.
\newblock Formal model and policy specification of usage control.
\newblock {\em ACM Trans. Inf. Syst. Secur.}, 8(4):351--387, 2005.

\bibitem{Zhang-2004-p1-10}
Xinwen Zhang, Jaehong Park, Francesco Parisi-Presicce, and Ravi Sandhu.
\newblock A logical specification for usage control.
\newblock In {\em SACMAT '04: Proceedings of the ninth ACM symposium on Access
  control models and technologies}, pages 1--10, New York, NY, USA, 2004. ACM.

\bibitem{Zhao03systemdependence}
Jianjun Zhao and Martin Rinard.
\newblock System dependence graph construction for aspect-oriented programs.
\newblock Technical Report MIT-LCS-TR-891, Laboratory for Computer
  Science.MIT., 2003.

\end{thebibliography}

%
%
%
%





\end{document}